# Optimization of Sound Energy Reduction in the Polycarbonate Plate Reinforced with Carbon Nanotubes

*Parinus Vedadi*[(a)]*, *Edris* Faizabadi[(b)]

(a) *School of Physics, Iran University of Science and Technology, 16846 Tehran, Iran, parinusv@gmail.com*

(b) *School of Physics, Iran University of Science and Technology, 16846 Tehran, Iran,* edris@iust.ac.ir

- *Corresponding Author*

## Abstract

In order to develop a new method for sound insulation materials reinforcement, in this study, the effect of carbon nanotubes on polycarbonate plates as a sample of materials used in acoustic insulators, have been investigated via numerical methods including finite element method (FEM) and finite difference time domain method (FDTDM). It was observed that the addition of carbon nanotubes to the polycarbonate material at frequencies below 1000 Hz increased the reflection of the acoustic wave from the surface of the polycarbonate plate and also decreased the amplitude of the acoustic wave pressure within the material. Therefore, it can be concluded that the reinforcement of polycarbonate plates by carbon nanotubes can reduce the intensity and help absorption of acoustic energy in the plate, and as a result, reduce the transmitted acoustic energy through the plate at low frequencies. Thereupon, these reinforced polycarbonate plates can be used in situations where a sound-absorbing material is needed next to the reflector material in the sound barriers.

## 1. Introduction

The development of industries and the expansion of cities cause noise pollution to become more prominent, which intensifies the need to consider its harms and disadvantages. In order to control and reduce noise pollution, it is important to identify the sources of sound and select the appropriate method for noise reduction. Using sound absorber materials in the sound propagation path is a common method to reduce unwanted sounds. Sound-absorbing materials are sometimes used individually and sometimes in combination with other materials, forming composite absorbs to minimize the sound transmission. In recent years, we have witnessed an increase in the use of nanomaterials in reinforcing acoustic absorbers; the purpose of the present study is to investigate the use of carbon nanotubes in acoustic absorbers and their optimization via numerical modeling.

According to the studies related to the use of nanomaterials in acoustics, Tang and Yan have used nano-fibrous material for low-frequency noise reduction and to improve the acoustic properties of sound absorbers. They observed that the fibrous material enhances noise reduction efficiency at low frequencies (Tang and Yan, 2017). Zhang et al. used fluid-conveying carbon nanotubes to absorb acoustic nano-wave in a determined frequency range by adjusting the length of the nanotubes and the liquid velocity (Mechanica *et al.*, 2016). Kudelcik et al. used acoustic spectroscopy to study the MWCNT functionalized by $Fe_3O_4$ that broke down in transformer oil affected by a magnetic field (Kúdel *et al.*, 2020). Ivanov et al. investigated the orientation of cylindrical nano objects in an acoustic tube and their effects, using the absorption spectrum. They could experimentally achieve the effect of the ratio and dimension of cylindrical nano objects and nano fibers in sound reduction using the absorption spectroscopy method (Ivanov *et al.*, 2017). Liu et al. have investigated the effect of nanotubes on acoustic wave absorption in polymers (Liu *et al.*, 2018). Loshkarev et al. investigated aqueous dispersions of different orientations of carbon nanotubes by calculating the ultrasound attenuation. They concluded that the perpendicular orientation is more effective in sound attenuation in frequencies higher than 15 MHz, and this effect increases as frequency increases (Loshkarev *et al.*, 2016). Wu et al. studied the capability of graphene foam /carbon nanotube /polydimethylsiloxane composites in damping of low-frequency sound waves and achieved the absorption coefficient of 0.3 in the frequency range of 100 to 1000 Hz. Besides, they determined proper thickness and porosity to achieve the maximum absorption in different frequency bands. They concluded that the achieved composition has the maximum absorption in low frequencies (Wu *et al.*, 2017). Qian et al. grew super-aligned carbon nanotubes (SACNT) on micro perforated panels (MPP) to improve the acoustic performance of panels and concluded that these panels have better acoustic performance in low frequencies, and the absorption depends on the length of the SACNTs (Qian *et al.*, 2014). Orfali added 12 different amounts of CNT and Silicon Oxide Nano-powder to the Polyurethane foam experimentally. The acoustic absorption of the 12 samples was investigated, and the sound transmission loss was measured up to 80dB (Orfali, 2015). Ayub et al. investigated the acoustic performance of absorbers that are made up of vertically aligned CNT forests experimentally, using two-microphone methods in an impedance tube. They concluded that the forests with 3mm thick CNTs could achieve acoustic absorption of 10% in the frequency range of 125 Hz to 4KHz and the acoustic absorption of a combined composite of CNT forest and a conventional acoustic material can be increased up to 5-10% (Ayub *et al.*, 2017). Nie et al. investigated the effective modulus of elasticity and Poisson's ratio of a Hybrid Aluminum Matrix Composite (HAMC) that was reinforced with ceramic particles and Carbon Nanotubes (CNTs) (Nie, Wang and He, 2020).

According to Ayub, the addition of carbon nanotubes to another material could affect the absorption of sound waves more than pure carbon nanotubes (CNT Forest) (Ayub *et al.*, 2017) And according to (Orfali, 2015)Adding the CNTs to the composition of another sound-absorbing material can improve the sound transmission loss. Therefore, further to the study by Orfali and Ayub et al. in this study, the sound wave energy loss is investigated while different amounts of CNTs are added to the acoustic media. Besides, to investigate the effect of CNTs on acoustic performance, solid polycarbonate is used instead of polyurethane as the host material, which has an acoustic reflection property along with an absorbing property. For this purpose, in order to control the accuracy of the numerical model, first, the experimental CNT model made by Ayub and his colleagues was modeled with the finite element model, and the absorption coefficient of the theoretical and experimental results for the CNT vertically facing the sound wave are compared. Then, the CNTs with random orientation are added to the host material, which is a polycarbonate sheet, and the effect of different amounts of carbon nanotubes on the sound behavior of the polycarbonate sheet is investigated.

According to Ayub, the addition of carbon nanotubes (CNTs) to another material could affect the absorption of sound waves more than pure carbon nanotubes (CNT Forest) (Ayub *et al.*, 2017). In agreement, recent studies have shown that the structural arrangement and orientation of CNTs, especially when functionalized or combined with other materials like $Fe_3O_4$ magnetic nanoparticles, significantly influence acoustic attenuation due to anisotropic interactions and chain-like formations under external fields (Kúdel *et al.*, 2020). According to (Orfali, 2015), adding CNTs to the composition of another sound-absorbing material can improve the sound transmission loss. Similarly, (Arunkumar *et al.*, 2021) demonstrated that functionally graded CNT-reinforced polymer composites enhance vibro-acoustic performance due to improved stiffness and tailored material properties. Therefore, further to the study by (Orfali, 2015; Ayub *et al.*, 2017), as well as more recent works (Kúdel *et al.*, 2020; Arunkumar *et al.*, 2021; Li *et al.*, 2025), in this study, the sound wave energy loss is investigated while different amounts of CNTs are added to the acoustic media. Besides, to investigate the effect of CNTs on acoustic performance, solid polycarbonate is used instead of polyurethane as the host material, which has an acoustic reflection property along with an absorbing property. For this purpose, in order to control the accuracy of the numerical model, first, the experimental CNT model made by Ayub and his colleagues was modeled with the finite element model, and the absorption coefficient of the theoretical and experimental results for the CNT vertically facing the sound wave are compared. Then, the CNTs with random orientation are added to the host material, which is a polycarbonate sheet, and the effect of different amounts of carbon nanotubes on the sound behavior of the polycarbonate sheet is investigated.

## 2. Methodology and Modelling

### *2-1. Numerical implementation*

Ayub et al. has measured the acoustic absorption coefficient of the CNT forest in an impedance tube using two microphones. A 3 mm length CNT forest was grown vertically on a silicon wafer substrate and the CNT samples were mounted according to the configurations showed in figure-1. The tested specimen is placed in a sample holder at the end of a tube and a plane wave is generated by a sound source that propagates the wave through the sample.

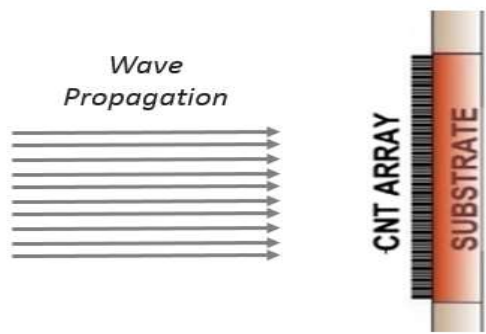

Figure 1- the CNT sample configuration towards the acoustic source (M.Ayub)

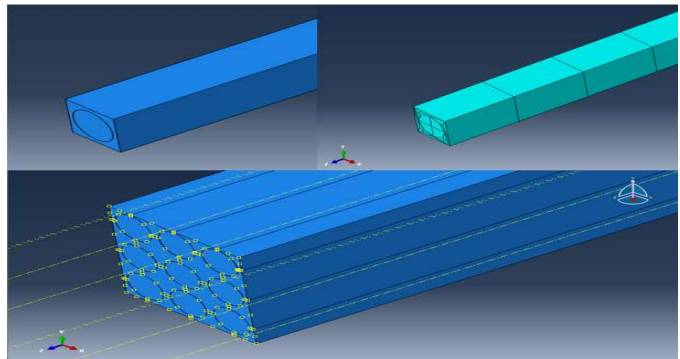

Figure 2- Finite element model of carbon nanotubes surrounded by air

In the continuation of the mentioned experimental model, multi-walled CNTs with the same specifications as the experimental model were individually modeled in the finite element software ABAQUS. The reason for the single modeling of the nanotube and not modeling the whole set of nanotubes is the increase in time and calculation in the finite element modeling process. For this purpose, several nanotubes were modeled, and the sound wave was applied to the sample, and after examining the behavior of different nanotubes and observing their similar behavior, to simplify the model and speed up the finite element calculation, a single nanotube was modeled in the face of sound wave and its behavior was examined at different frequencies.

In this model, the inner diameter of the nanotubes (0.34 nm) was neglected because it was very small compared to the sample length. Outer radius of 35 nm, density of 1.6 g/cm$^3$, modulus of elasticity of 1 TPa, (David A. Bies, 2009), 3 mm length, sweep technique for meshing, hexagonal mesh type, AC3D8 element type, and the density of 1.225 kg/m$^3$ and modulus of elasticity of 101 KPa for the surrounding air were defined. Then, the absorption coefficient obtained at different frequencies was compared to those of Ayub (Ayub *et al.*, 2017) In Figure 3.

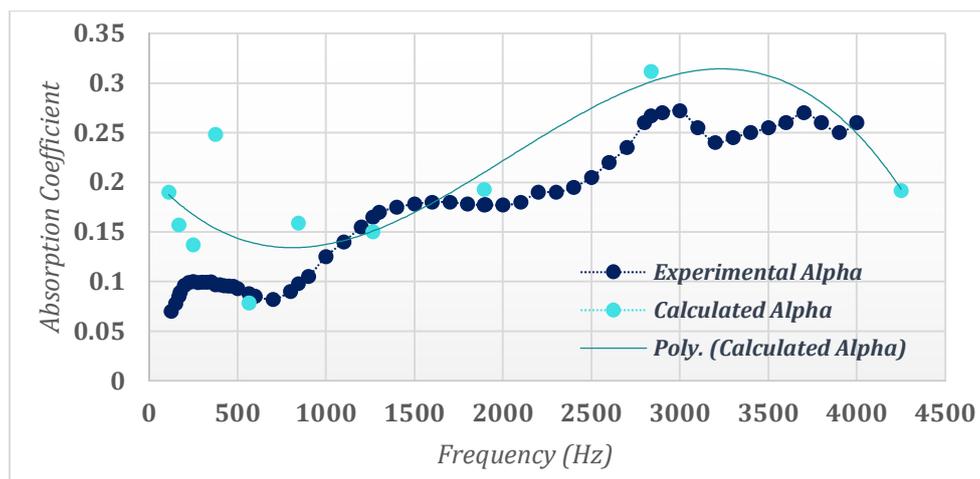

Figure 3- comparing the absorption coefficient of Finite Element model and the experimental model of (Ayub et al., 2017)

Figure 3 shows the absorption coefficient of the numerical and experimental models. The experimental absorption coefficient was obtained by (Ayub *et al.*, 2017) that made the CNT forest and measured its absorption coefficient in an impedance tube. The numerical absorption coefficient (Calculated alpha) shown in Figure 3 was obtained using the finite element model, which is described in the previous paragraphs.

It is observed that the difference between experimental and theoretical values is higher at frequencies lower than 400 Hz, and at higher frequencies, the difference gets lower. The reason for the difference between the two sets of data can be noted as follows. In the experimental model, a circular set of multi-walled carbon nanotubes with a length of 3 mm and a diameter of 2 cm in the impedance tube is exposed to the sound wave, and its absorption coefficient is calculated. In the theoretical model, which is modeled in finite element software ABAQUS, the ratio of diameter to length of carbon nanotubes is very small. This causes a huge increase in the number of elements, which increases the time and calculation process.

In the finite element model, firstly, 9 nanotubes were modeled next to each other, and similar to the experimental model, they were exposed to the sound wave with orientation parallel to the direction of sound wave propagation. After observing the similar behavior of nanotubes placed in different positions inside the 9-element pack, in the next step, a single nanotube was modeled and exposed to the sound wave, and its absorption coefficient was calculated at different frequencies. The reason for the difference between the results of the theoretical and experimental models can be attributed to the lack of modeling of carbon nanotube sets in the finite element model.

Furthermore, the difference observed between the numerical and experimental results shows the accuracy of the finite element model, and it is observed that this difference increases at low frequencies. So, it is concluded that to achieve the appropriate accuracy of the finite element model at low frequencies, the set of carbon nanotubes should be modeled instead of individual carbon nanotubes.

For this purpose, in the continuation of this research, instead of a single nanotube, the set of nanotubes in reinforcement of the acoustic panel was modeled to investigate the sound propagation behavior inside the panel.

### 2-2. Sound wave propagation Modelling

First, the sound wave propagation inside the polycarbonate plate reinforced with CNTs was modelled by Time Difference Time Domain (FDTD) method according to (Causon and Mingham, 2010). Then in order to validate the FDTD model, a Finite Element (FE) model was used by ABAQUS Software and in the next step, the results of the two models were compared. The effective parameters on sound propagation are the effective density and the effective modulus of elasticity of the sound propagation media which depend on the amount of CNTs. Therefore, the mentioned parameters were calculated for different amounts of CNT inside the sound propagation media, and the propagation of sound was studied through the media.

Given that at the microscopic scale, the sound propagation study is difficult in porous media due to the geometrical complexities, considering the average values of the parameters involved in the acoustic field is sufficient (Jean Allard, 2009). The averaging should be done on a macroscopic scale, and the porosity dimensions should be sufficiently small for the acoustic wavelength and the homogenization medium (Jean Allard, 2009).

Therefore, given that the porosity dimensions in the medium, which are carbon nanotubes, are about 9 times smaller than the sample length (the sample size is 0.6 m, and the length of the nanotubes is 1 nm), homogenization of the medium was performed on the density and the modulus of elasticity. For this purpose, the volumetric equivalent method presented by (Rivin, 1999) was used. The final equation Rivin presented for calculating the system equivalent parameters, such as modulus of elasticity and density in the 'y' direction, is as equation (1).

$$\tilde{C}_{ijkl} = \frac{1}{Y}\int_Y \left(C_{ijkl}(y) + C_{ijmn}(y)\frac{\partial N_m^{kl}(y)}{\partial y_n}\right) dv \tag{1}$$

Where $\tilde{C}_{ijkl}$ is the equivalent mechanical parameter of the total media containing the host material and the guest material. $C_{ijkl}(y)$ is the mechanical parameter of the host material, $C_{ijmn}(y)$ is the mechanical parameter of the guest material that is periodically distributed along the y-direction of the host material with the period of Y, and $N_m$ is the distribution function, which is assumed uniform.

Therefore, the equivalent parameters of the polycarbonate plate reinforced with different amounts of CNT were calculated using the material properties presented in Table 1 and Equation 1.

*Table 1- The material properties used in simulations* (Demczyk et al., 2002; David A. Bies, 2009)

| Material | Modulus of Elasticity (GPa) | Density ($Kg/m^3$) |
|---|---|---|
| Polycarbonate | 2 | 1200 |
| Carbon Nanotubes | 1000 | 1360 |

The reason for choosing the polycarbonate material as the host material was that this material is mainly selected as a reflector of sound waves in the sound barriers. To improve the sound absorption of this material besides improving its reflecting feature, it can be useful to reduce the transmitted wave via two mechanisms of absorption and reflection.

The sample is simulated in ABAQUS finite element software in standard form. The geometry is defined as a plate with dimensions of $60 \times 60 \times 5\ cm$ (Figure 4), The boundary condition is defined as symmetrical on one side of the X direction and both sides of the Y direction, so that the panel's continuity is considered. The load is defined as the acoustic pressure that is applied from the side of the sample in the X-direction instantaneously. The FDTD model is simulated by a mathematical method according to (Causon and Mingham, 2010) in MATLAB software, with the geometrical features as follows:

$\{x.y \in [-10\ \pi. 10\ \pi]. z \in [-0.8\ \pi. 0.8\ \pi]\}$

The material parameters are considered using Table 1 and Equation 1.

The results of the two methods are then compared for verification and precision.

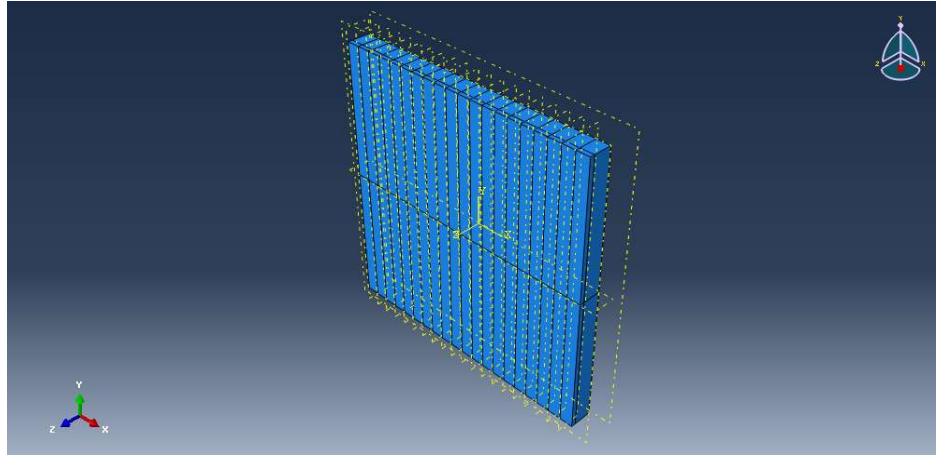

*Figure 4- Finite element model of the polycarbonate plate in ABAQUS Software*

Based on the method developed in the previous section, the sound pressure in two random frequencies through the sample in both models is presented as follows. Figures 5 and 6 represent the sound pressure curves through the reinforced polycarbonate plate with 0.1% CNT at the frequency 4620 Hz. It is observed that in both figures, 1.5 periods exist with pressure amplitudes of about 1 Pa. It should be noted that the only difference between the two samples is in geometry. The length of the sample was chosen as 0.6 meters in the FE model, and $20\,\pi$ in the FDTD model.

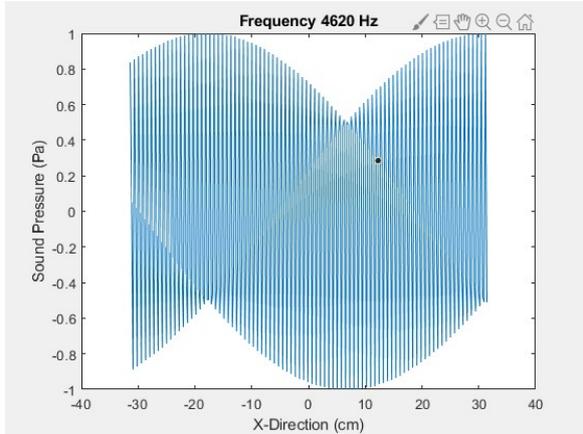
Figure 6- Sound pressure diagram along the sample at frequency 4620 Hz and CNT value of 0.1% using FDTD model

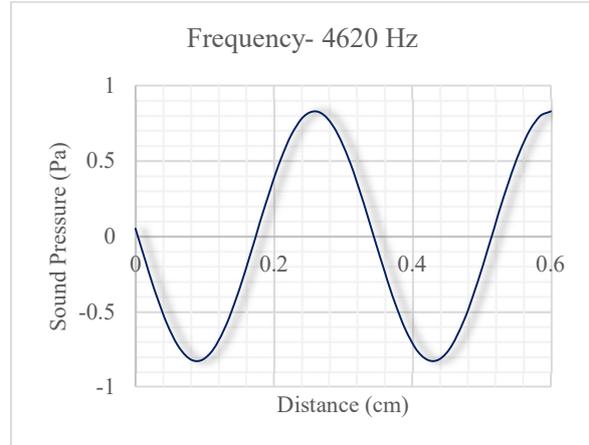
Figure 5- Sound pressure diagram along the sample at frequency 4620 Hz and CNT value of 0.1% using FE model

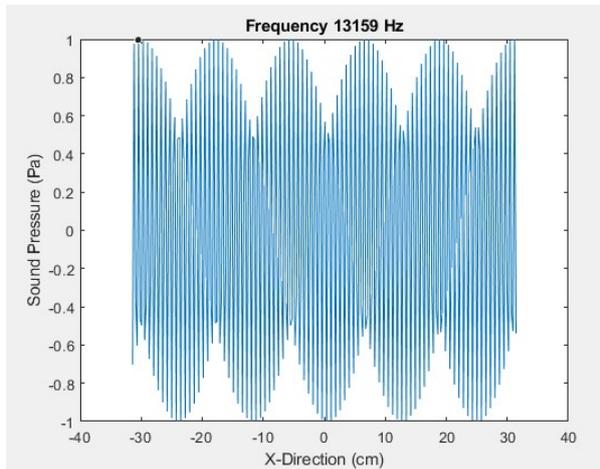
Figure 8- Sound pressure diagram along the sample at frequency 13159 Hz and CNT value of 0.1% using FDTD model

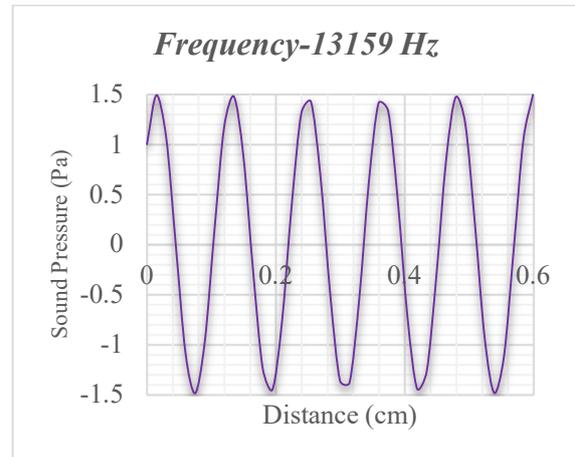
Figure 7- Sound pressure diagram along the sample at frequency 13159 Hz and CNT value of 0.1% using FE model

Figures 7 and 8 represent the sound pressure curves through the reinforced polycarbonate plate with 0.1% CNT at a frequency of 13159 Hz. It is observed that in both figures; 5 periods exist with pressure amplitudes of about 1-1.5 Pa that provide acceptable differences.

### 3. Results and discussion

The effect of adding 0.1% of CNTs to the polycarbonate plate was investigated, and the results are presented as follows. There are two samples compared. Both samples had the same geometry, loading, and boundary conditions, and the only difference was in their material properties, which are dependent on the amount of CNT. One sample was pure polycarbonate (No CNT), and the second one was reinforced with 0.1% CNT.

In Figure 9, the sound intensity in both samples at 20 Hz is presented. It is observed that despite the identical input of the acoustic wave in both samples, the acoustic wave intensity upon arrival of the sample with 0.1% CNT is about 50% less than the pure sample. The reason for this phenomenon is that by adding carbon nanotubes to the sample, the sample becomes denser and has more modulus of elasticity, which corresponds to the increase in the density of matter particles, thereby increasing the possibility of sound wave reflection from the material surface. The approximately similar result can be found in figures 9(b) to 9(g).

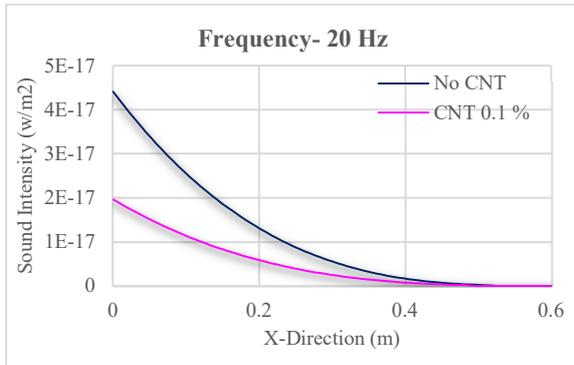

*(a) 20 Hz*

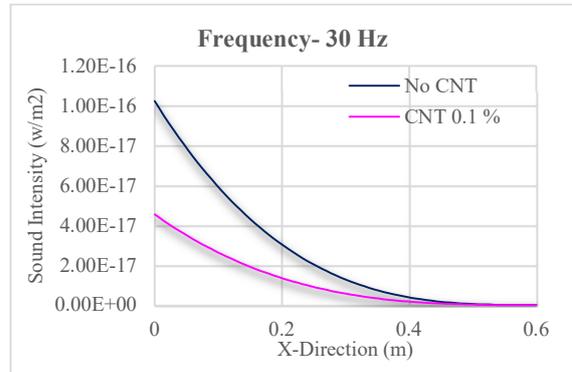

*(b) 30 Hz*

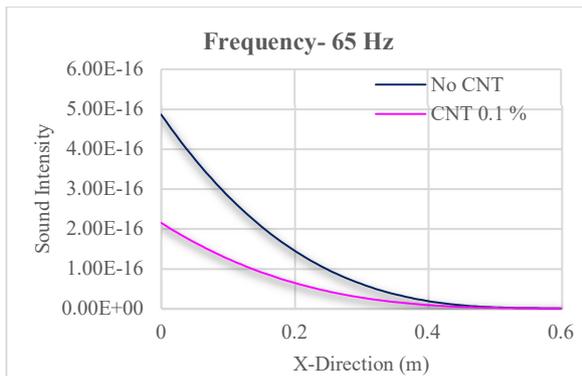

*(c) 65 Hz*

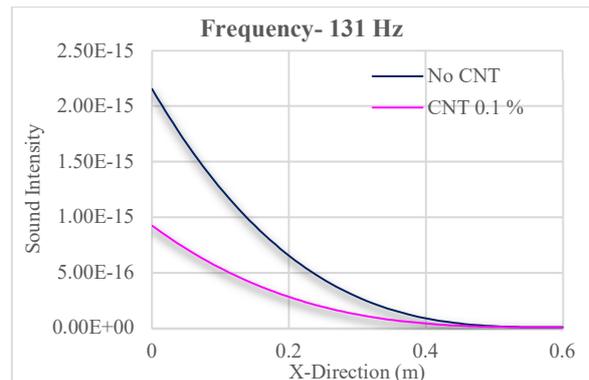

*(d) 131 Hz*

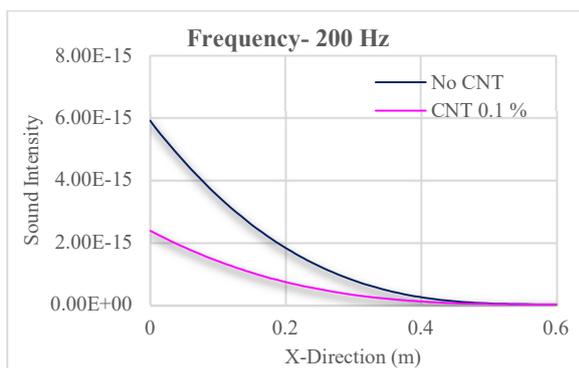

*(e) 200 Hz*

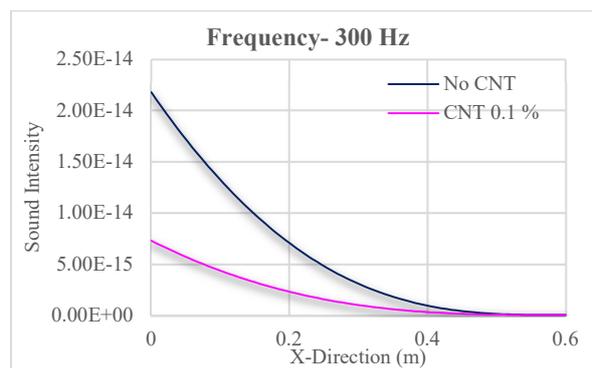

*(f) 300 Hz*

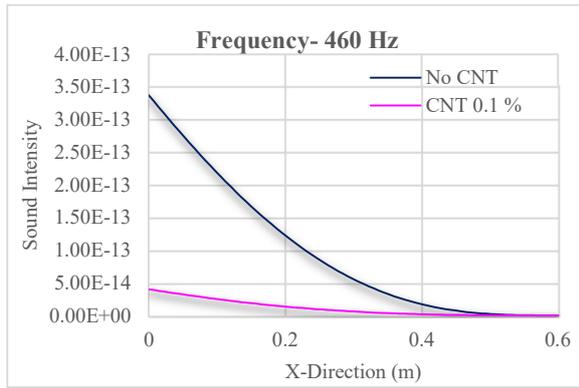

*(g) 460 Hz*

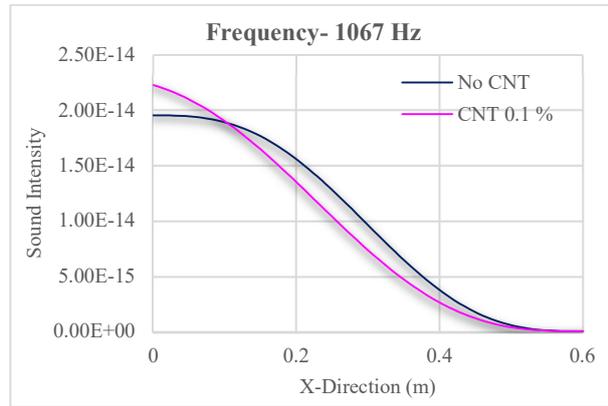

*(h) 1067 Hz*

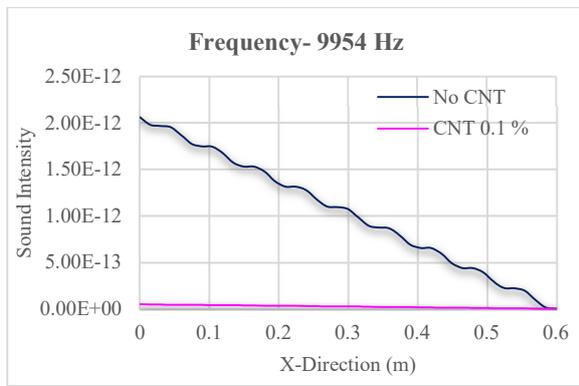

*(i) 9954 Hz*

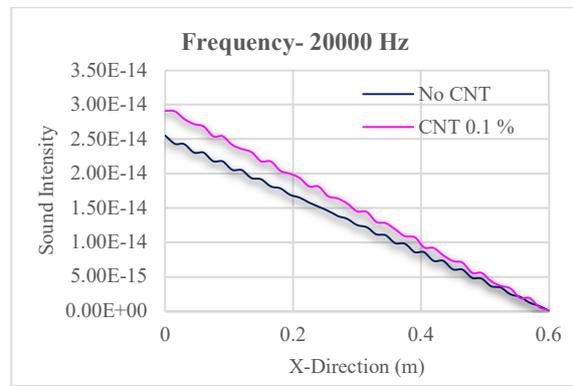

*(j) 20000 Hz*

*Figure 9- Comparing soundwave intensity in pure polycarbonate vs. polycarbonate reinforced with 0.1% CNT at different frequencies*

In Figures 9(b) to 9(g), the soundwave intensities through the pure polycarbonate and polycarbonate reinforced with 0.1% CNT are presented for frequencies 30, 65, 131, 200, 300, and 460 Hz, respectively. It can be seen that the soundwave intensity upon entry of the reinforced sample is about 60-87% lower than that of the pure sample. In Figure 9(i), the soundwave intensifies through pure polycarbonate and 0.1% carbon nanotube reinforced polycarbonate samples are presented at 9954 Hz. It can be seen that the soundwave intensity upon arrival of the reinforced sample is about 0.002 times the soundwave intensity of the pure sample. In Figure 16, the soundwave intensities through pure polycarbonate and 0.1% carbon nanotube reinforced polycarbonate samples are presented at 1067 Hz, showing a variable behavior of sound intensity. In Figure 9(j), the soundwave intensities through the two mentioned samples are presented at 20000 Hz. It can be seen that the soundwave intensity upon arrival of the reinforced sample is 1.14 times bigger than that of the pure sample.

In summary, it can be stated that according to the study of soundwave intensity through two samples of pure polycarbonate plate and polycarbonate plate reinforced with 0.1% carbon nanotube at the frequency range 20 to 20,000 Hz (human hearing frequency range), It was observed that at some frequencies the sound intensity in the carbon nanotube-reinforced sample was lower than that of the pure sample, and at some frequencies the opposite was observed.

The main reason for this can be stated based on frequency analysis performed on pure polycarbonate plates and polycarbonate plates reinforced with carbon nanotubes; the majority of the natural frequencies of the sheets were above 1000 Hz. Given that most of the natural frequencies are very close together in the reinforced plates, the plates experience resonance at most of the frequencies above 1000 Hz. The results are such that no distinct sequence or pattern can be obtained from the graphs' analysis.

The above results indicate that although the input acoustic waves are identical in the two samples, the soundwave intensity of both samples at the beginning of the sample entry is different at all frequencies. The reason for these differences can be described by the sound wave reflection from the surface of the samples and the difference between their reflection indices. Since the energy (represented by sound intensity) carried by the input sound wave is proportional to the inverse of the impedance [14], the impedance of the media is calculated in both states of pure polycarbonate and reinforced polycarbonate plates. By definition, acoustic impedance is equal to the ratio of acoustic pressure to the particle velocity [14]. Accordingly, the impedance value is calculated by 1 Pascal pressure at the entrance surface of both media in the frequency range of 20-20000 Hz represented in Table 2. In Table 2, *v* and *Z* are the particle velocity and the calculated impedance values, respectively.

Table 2- Calculated impedance values of pure polycarbonate (No CNT) plate and polycarbonate plate reinforced with 0.1% CNT at different frequencies

| Frequency (Hz) | No CNT | | 0.1% CNT | |
|---|---|---|---|---|
| | v | Z | v | Z |
| 20 | $87.38 \times 10^{-18}$ | $1.14 \times 10^{16}$ | $38.85 \times 10^{-18}$ | $2.57 \times 10^{16}$ |
| 30 | $202.6 \times 10^{-18}$ | $4.93 \times 10^{15}$ | $89.97 \times 10^{-18}$ | $11.11 \times 10^{15}$ |
| 65 | $962.4 \times 10^{-18}$ | $1.04 \times 10^{15}$ | $424.1 \times 10^{-18}$ | $2.35 \times 10^{15}$ |
| 130 | $4.26 \times 10^{-15}$ | $2.35 \times 10^{14}$ | $1.82 \times 10^{-15}$ | $5.49 \times 10^{14}$ |
| 200 | $11.7 \times 10^{-15}$ | $8.54 \times 10^{13}$ | $4.7 \times 10^{-15}$ | $21.27 \times 10^{13}$ |
| 300 | $43.18 \times 10^{-15}$ | $2.31 \times 10^{13}$ | $14.4 \times 10^{-15}$ | $6.94 \times 10^{13}$ |
| 460 | $668.7 \times 10^{-15}$ | $1.49 \times 10^{12}$ | $79.65 \times 10^{-15}$ | $12.55 \times 10^{12}$ |
| 750 | $134.87 \times 10^{-15}$ | $7.41 \times 10^{12}$ | $574.5 \times 10^{-15}$ | $1.74 \times 10^{12}$ |
| 1067 | $37.76 \times 10^{-15}$ | $2.64 \times 10^{13}$ | $43.86 \times 10^{-15}$ | $2.28 \times 10^{13}$ |
| 9954 | $3.82 \times 10^{-12}$ | $2.61 \times 10^{11}$ | $47.14 \times 10^{-15}$ | $212.13 \times 10^{11}$ |
| 20000 | $57.1 \times 10^{-15}$ | $1.754 \times 10^{13}$ | $14.1 \times 10^{-15}$ | $7.1 \times 10^{13}$ |

In order to control the effect of the impedance on the soundwave reflection, the impedance ratio and the sound intensity ratio of the two samples are calculated in different frequencies and represented in table 3.

Table 3- Impedance and the sound intensity ratios of the two samples at different frequencies

| Frequency (Hz) | Impedance Ratio (No CNT Sample/ 0.1% CNT Sample) | Intensity Ratio (No CNT Sample/ 0.1 % CNT Sample) | | Difference Between Impedance Ratio and Intensity Reverse Ratio |
|---|---|---|---|---|
| | | Ratio | 1/Ratio | |
| 20 | 0.44358 | 2.249046593 | 0.444633 | **0.001581166** |
| 30 | 0.443744 | 2.251724942 | 0.444104 | **0.000539828** |
| 65 | 0.442553 | 2.269277954 | 0.440669 | **0.002832616** |
| 130 | 0.428051 | 2.342040457 | 0.426978 | **0.00163987** |
| 200 | 0.401504 | 2.490607134 | 0.401509 | **6.4072E-06** |
| 300 | 0.332853 | 2.996971146 | 0.33367 | **0.00141643** |
| 460 | 0.118725 | 8.401923785 | 0.11902 | **0.00085691** |
| 750 | 4.258621 | 0.235224945 | 4.25125 | **0.003571787** |
| 1067 | 1.157895 | 0.876491253 | 1.140913 | **0.015781755** |
| 9954 | 0.012304 | 42.04215398 | 0.023786 | **0.103512693** |
| 20000 | 0.247042 | 0.875727862 | 1.141907 | **1.800412018** |

It is observed that adding CNTs to the polycarbonate plate increases the plate impedance against acoustic waves and decreases the intensity of the entering sound wave in the plate at all frequencies except 20000 Hz. Therefore, it can be stated that increasing the sample impedance increases the reflectance of the sound wave from its surface at low frequencies, which causes a difference between the sound intensity entering the samples despite the identical characteristics of the applied sound pressure to both samples.

Then, by studying the propagation of sound pressure inside samples with different amounts of carbon nanotubes, the effect of the amount of carbon nanotubes on the sound pressure applied to the samples is investigated.

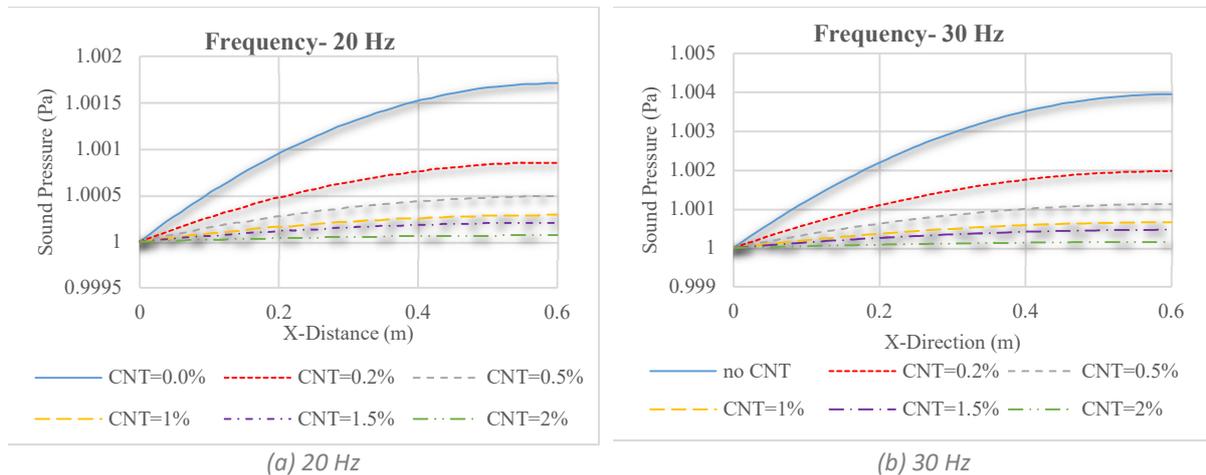

*(a) 20 Hz*  *(b) 30 Hz*

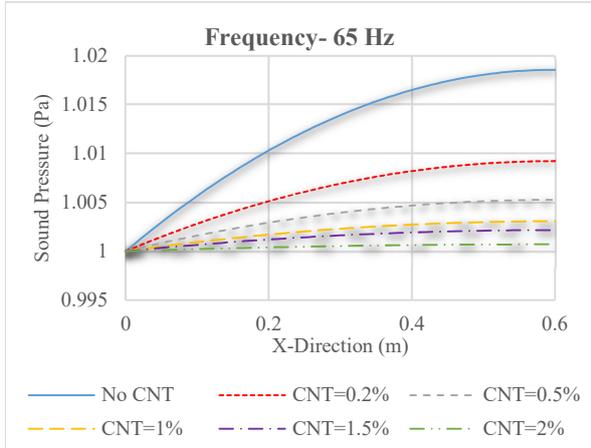
*(c) 65 Hz*

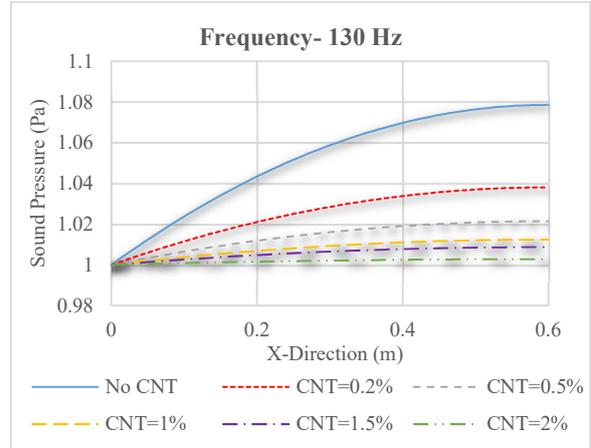
*(d) 130 Hz*

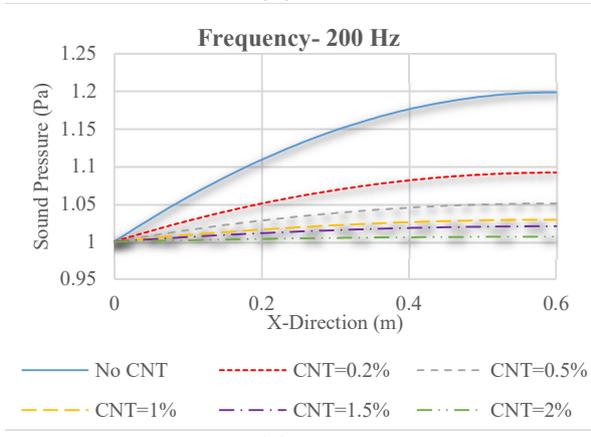
*(e) 200 Hz*

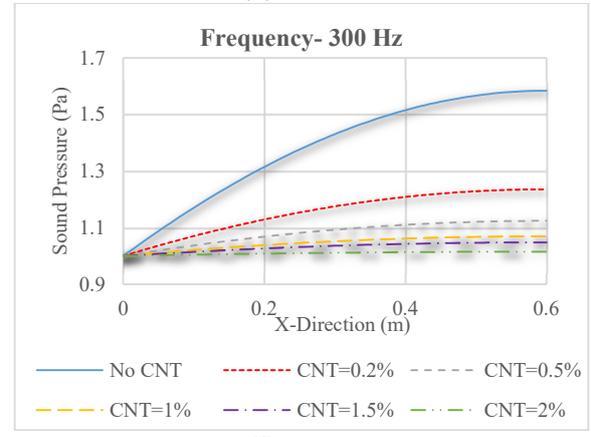
*(f) 300 Hz*

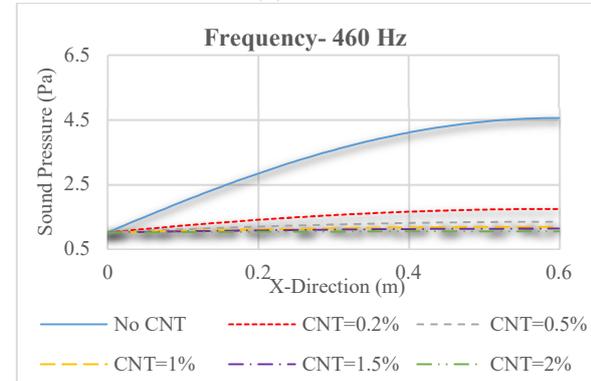
*(g) 460 Hz*

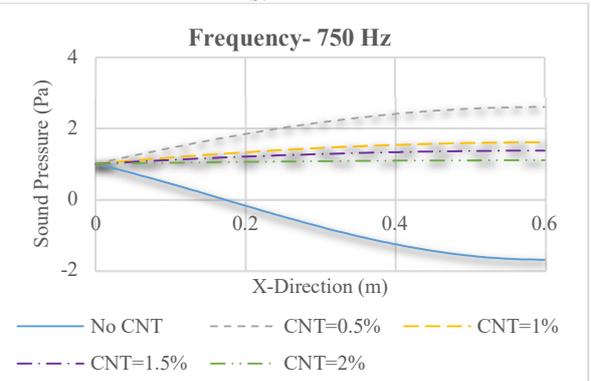
*(h) 750 Hz*

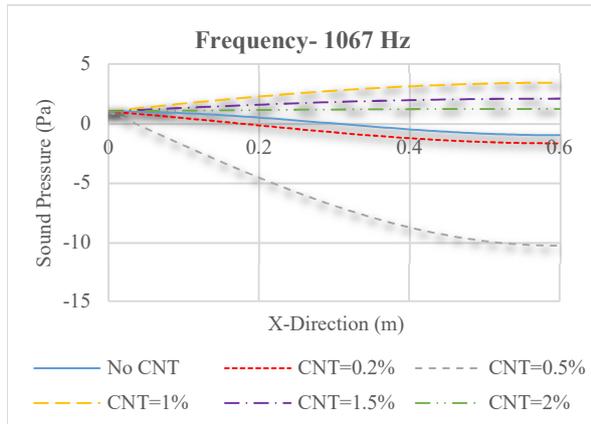
*(i) 1067 Hz*

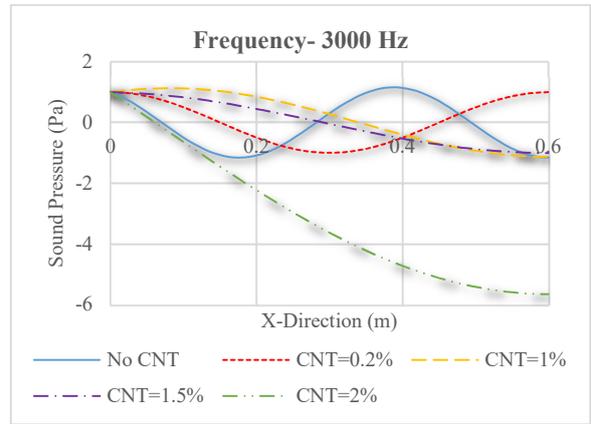
*(j) 3000 Hz*

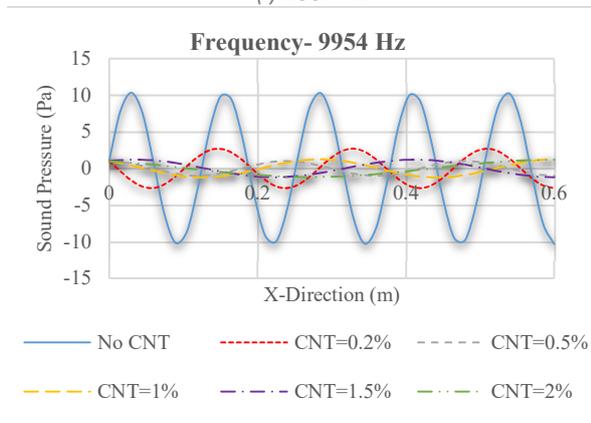
*(k) 9954 Hz*

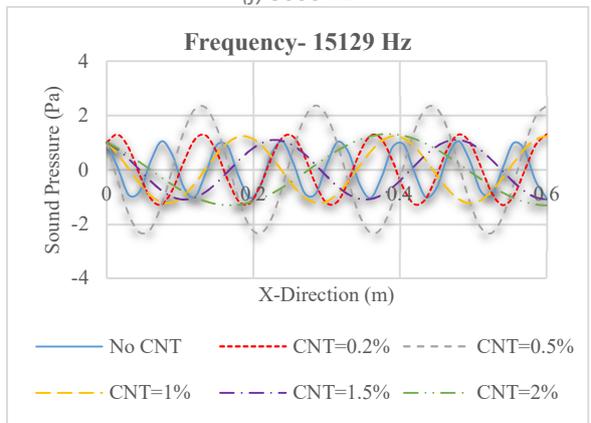
*(l) 15129 Hz*

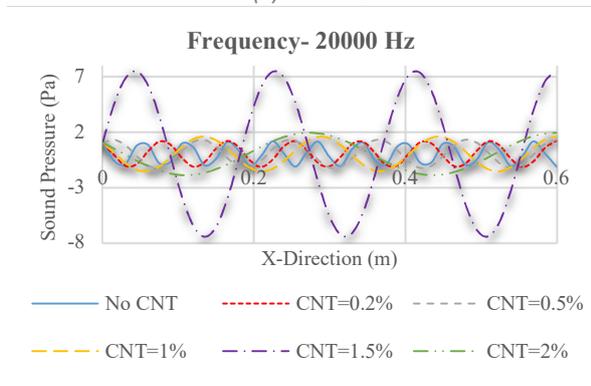
*(m) 20000 Hz*

*Figure 10- Sound pressure inside the samples of polycarbonate with different amounts of CNT added at different frequencies*

In Figure 10(a), it can be observed that at 20 Hz, the pure sample has the highest sound pressure amplitude, and the sound pressure amplitude decreases as the carbon nanotubes increase. In Figure 10(b) at 30 Hz, the pure sample has the highest sound pressure domain, and the sound pressure decreases as the CNT value increases. Besides, comparing figures 10(a) and 10(b) gives the result that the sound pressure domain at 30 Hz is higher than the one at 20 Hz. The same results are observed in figures 10(c) to 10(g). In figure 10(h) at 750 Hz, the pure sample has the highest pressure domain and the CNT value increases, the sound pressure domain decreases except the diagram belong to 0.2% CNT which is drawn out of the figure due to its high pressure domain which is caused by the resonance occurred at the plate according to the frequency analysis performed on the plate with 0.2% amount of CNT.

In figures 10(i), 10(j), 10(l), and 10(m), it is observed that the sound pressure domains are not proportional to the amount of CNTs added to the plates and there are different behaviors observed at frequencies higher than 1000 Hz. The main reason for this is that based on frequency analysis, in pure polycarbonate plates and polycarbonate plates reinforced with carbon nanotubes, the majority of the natural frequencies of the plates were above 1000 Hz and this resulted in resonances occurring in plates at very close frequencies above 1000 Hz leading to complex behaviors with no regular pattern.

Therefore, for a proper design of a polycarbonate plate reinforced with CNTs is to choose the value of CNTs added to the plate is chosen at the first step, then, having the plate dimensions, the frequency analysis is performed, and the eigen-frequencies are determined. Then, comparing the eigen-frequencies of the panel with the dominant frequencies of the soundwave desired to be reduced can help to choose if the selected percent of CNTs is proper when these two sets of frequencies have the lowest values.

Besides, the obtained graphs denote that the thickness of the polycarbonate plate in front of the acoustic wave (shown in the graphs in the X direction) has an effective role in decreasing the acoustic intensity, and increasing the plate thickness causes a significant decrease in the intensity of the sound wave.

However, due to the fact that in industrial applications the increase of plate thickness is not simply done, it is possible to reduce the amplitude of the acoustic wave by the use of increase in reflection, decrease in intensity and amplitude of the transmitted sound wave to increase the acoustic performance of polycarbonate panels, and increasing thickness can be used as a final choice in applications with high pressure sound.

## 4. Conclusion

In order to investigate the effect of adding carbon nanotubes to sound absorber materials, polycarbonate was used as the host material because of its general use in sound barriers. Then, the changes of the mechanical properties of the polycarbonate material, including density and modulus of elasticity, were determined by adding carbon nanotubes with different values (from 0.1% to 2% of the volume). Then, by FE and FDTD models of polycarbonate panel with certain mechanical properties, the sound wave with initial amplitude of 1 Pascal was applied to the samples and their behavior against the sound wave was studied. First, observing the sound intensity reduction by adding 0.1% CNT to the host material showed the increase in sound wave reflection from the sample surface due to adding CNTs which is proved by calculating the change in the impedance of the material. Further, by investigating the sound propagation behavior it was observed that the addition of carbon nanotubes to the polycarbonate plate reduces the sound pressure domain and intensity transferred from the polycarbonate plate, especially at frequencies below 1000 Hz. Therefore, it can be concluded that the reinforcement of polycarbonate plates by carbon nanotubes can reduce the intensity and help absorption of acoustic energy in the plate, and as a result, reduce the transmitted acoustic energy through the plate at low frequencies.

## 5. Acknowledgment

This work was supported by Iran University of Science and Technology (IUST) Grant no. 160/8756.